\newcommand{\CLASSINPUTtoptextmargin}{19mm}%
\newcommand{\CLASSINPUTbottomtextmargin}{43mm}%
\newcommand{\CLASSINPUTinnersidemargin}{12.9mm}%
\newcommand{\CLASSINPUToutersidemargin}{12.9mm}%
\documentclass[conference,10pt,a4paper]{IEEEtran}% requires IEEEtran V1.8+
\usepackage{cite}
\usepackage{microtype}
\usepackage{amsmath}
\usepackage{amssymb}
\usepackage{graphicx}
\usepackage{upgreek}
\usepackage{times}

\usepackage{xcolor}
\usepackage{multirow}
\usepackage{diagbox}

\usepackage[style=default]{caption}
\usepackage{subcaption}

\usepackage{soul}

\usepackage{pgfplots}
\pgfplotsset{compat=newest}
\pgfplotsset{plot coordinates/math parser=false}
\newlength\figureheight
\newlength\figurewidth

\makeatletter

\def\@maketitle{\newpage
\bgroup\par\addvspace{0.5\baselineskip}\centering%
\ifCLASSOPTIONtechnote% technotes
   {\bfseries\large\@IEEEcompsoconly{\sffamily}\@title\par}\vskip 1.3em{\lineskip .5em\@IEEEcompsoconly{\sffamily}\@author
   \@IEEEspecialpapernotice\par{\@IEEEcompsoconly{\vskip 1.5em\relax
   \@IEEEtitleabstractindextextbox{\@IEEEtitleabstractindextext}\par
   \hfill\@IEEEcompsocdiamondline\hfill\hbox{}\par}}}\relax
\else% not a technote
   \vskip0.2em{\EuMWtitlesize\ifCLASSOPTIONtransmag\bfseries\LARGE\fi\@IEEEcompsoconly{\sffamily}\@IEEEcompsocconfonly{\normalfont\normalsize\vskip 2\@IEEEnormalsizeunitybaselineskip
   \bfseries\Large}\@title\par}\vskip1.0em\par% CAUSAL PRODUCTIONS change on this line
   \ifCLASSOPTIONconference%
      {\@IEEEspecialpapernotice\mbox{}\vskip\@IEEEauthorblockconfadjspace%
       \mbox{}\hfill\begin{@IEEEauthorhalign}\@author\end{@IEEEauthorhalign}\hfill\mbox{}\par}\relax
   \else% peerreviewca, peerreview or journal
      \ifCLASSOPTIONpeerreviewca
         {\@IEEEcompsoconly{\sffamily}\@IEEEspecialpapernotice\mbox{}\vskip\@IEEEauthorblockconfadjspace%
          \mbox{}\hfill\begin{@IEEEauthorhalign}\@author\end{@IEEEauthorhalign}\hfill\mbox{}\par
          {\@IEEEcompsoconly{\vskip 1.5em\relax
           \@IEEEtitleabstractindextextbox{\@IEEEtitleabstractindextext}\par\hfill
           \@IEEEcompsocdiamondline\hfill\hbox{}\par}}}\relax
      \else% journal, peerreview or transmag
         \ifCLASSOPTIONtransmag
           {\@IEEEspecialpapernotice\mbox{}\vskip\@IEEEauthorblockconfadjspace%
            \mbox{}\hfill\begin{@IEEEauthorhalign}\@author\end{@IEEEauthorhalign}\hfill\mbox{}\par
           {\vspace{0.5\baselineskip}\relax\@IEEEtitleabstractindextextbox{\@IEEEtitleabstractindextext}\vspace{-1\baselineskip}\par}}\relax
         \else% journal or peerreview
           {\lineskip.5em\@IEEEcompsoconly{\sffamily}\sublargesize\@author\@IEEEspecialpapernotice\par
           {\@IEEEcompsoconly{\vskip 1.5em\relax
            \@IEEEtitleabstractindextextbox{\@IEEEtitleabstractindextext}\par\hfill
            \@IEEEcompsocdiamondline\hfill\hbox{}\par}}}\relax
         \fi
      \fi
   \fi
\fi\par\addvspace{0.0\baselineskip}\egroup}% CAUSAL PRODUCTIONS change on this line, reduce the vspace from 0.5\baselineskip to 0.0

\def\EuMWtitlesize{\@setfontsize{\EuMWtitlesize}{24}{24pt}}% CAUSAL PRODUCTIONS change on this line
\def\EuMWauthorsize{\@setfontsize{\EuMWauthorsize}{11}{11pt}}% CAUSAL PRODUCTIONS change on this line
\def\EuMWaffilsize{\@setfontsize{\EuMWaffilsize}{10}{10pt}}% CAUSAL PRODUCTIONS change on this line
\def\EuMWcaptionsize{\@setfontsize{\EuMWcaptionsize}{9}{10pt}}% CAUSAL PRODUCTIONS change on this line
\def\EuMWbibsize{\@setfontsize{\EuMWbibsize}{8}{10pt}}% CAUSAL PRODUCTIONS change on this line

\def\@IEEEauthorblockNstyle{\EuMWauthorsize\@IEEEcompsocnotconfonly{\sffamily}\@IEEEcompsocconfonly{\large}}%CAUSAL PRODUCTIONS removed sublargesize to get correct EuMWauthorsize
\def\@IEEEauthorblockAstyle{\EuMWaffilsize\@IEEEcompsocnotconfonly{\sffamily}\@IEEEcompsocconfonly{\itshape}\@IEEEcompsocconfonly{\large}}%CAUSAL PRODUCTIONS removed normalsize to get correct EuMWaffilsize
\def\@IEEEauthordefaulttextstyle{\EuMWauthorsize\@IEEEcompsocnotconfonly{\sffamily}\sublargesize}%CAUSAL PRODUCTIONS

\def\thebibliography#1{\section*{\refname}%
    \addcontentsline{toc}{section}{\refname}%
    \EuMWbibsize\@IEEEcompsocconfonly{\small}\vskip 0.3\baselineskip plus 0.1\baselineskip minus 0.1\baselineskip% CAUSAL PRODUCTIONS change on this line
    \list{\@biblabel{\@arabic\c@enumiv}}%
    {\settowidth\labelwidth{\@biblabel{#1}}%
    \leftmargin\labelwidth
    \advance\leftmargin\labelsep\relax
    \itemsep \IEEEbibitemsep\relax
    \usecounter{enumiv}%
    \let\p@enumiv\@empty
    \renewcommand\theenumiv{\@arabic\c@enumiv}}%
    \let\@IEEElatexbibitem\bibitem%
    \def\bibitem{\@IEEEbibitemprefix\@IEEElatexbibitem}%
\def\newblock{\hskip .11em plus .33em minus .07em}%
\ifCLASSOPTIONtechnote\sloppy\clubpenalty4000\widowpenalty4000\interlinepenalty100%
\else\sloppy\clubpenalty4000\widowpenalty4000\interlinepenalty500\fi%
    \sfcode`\.=1000\relax}

\long\def\@makecaption#1#2{%
\ifx\@captype\@IEEEtablestring%
\par\@IEEEtabletopskipstrut% strut used to align table caption with facing column
\else
\@IEEEfigurecaptionsepspace
\fi
\setbox\@tempboxa\hbox{\normalfont\footnotesize {#1.}\nobreakspace\nobreakspace #2}%
\ifdim \wd\@tempboxa >\hsize%
\setbox\@tempboxa\hbox{\normalfont\footnotesize {#1.}\nobreakspace\nobreakspace}%
\parbox[t]{\hsize}{\normalfont\footnotesize\noindent\unhbox\@tempboxa#2}%
\else
\ifCLASSOPTIONconference \hbox to\hsize{\normalfont\footnotesize\hfil\box\@tempboxa\hfil}%
\else \hbox to\hsize{\normalfont\footnotesize\box\@tempboxa\hfil}%
\fi\fi
\ifx\@captype\@IEEEtablestring%
\@IEEEtablecaptionsepspace
\else
\fi}

\newlength\tablecaptiontotableskip
\newlength\figuretocaptionskip
\def\@IEEEfigurecaptionsepspace{\vskip\figuretocaptionskip\relax}%
\def\@IEEEtablecaptionsepspace{\vskip\tablecaptiontotableskip\relax}%

\def\abstract{\normalfont%
\@IEEEabskeysecsize\bfseries\textit{\abstractname}\,\bfseries\textit{---}\,%
\@IEEEgobbleleadPARNLSP}%

\def\IEEEkeywords{\normalfont%
\@IEEEabskeysecsize\bfseries\textit{\IEEEkeywordsname}\,\bfseries\textit{---}\,%
\@IEEEgobbleleadPARNLSP}%
\def\endIEEEkeywords{\relax\vspace{0.67ex}%
\par\if@twocolumn\else\endquotation\fi%
\normalsize\normalfont}%

\def\@IEEEauthorblockNtopspace{0ex}
\def\@IEEEauthorblockAtopspace{1mm}
\def\tablename{Table}% EuMW: was TABLE in IEEETran, but we are now using capitalized caption text not smallcaps
\def\thetable{\arabic{table}}% EuMW: Tables are numbered using arabic not roman
\def\IEEEkeywordsname{Keywords}% use Keywords instead of Index Terms
\def\subsubsection{\@startsection{subsubsection}{3}{\z@}{1.5ex plus 1.5ex minus 0.5ex}%
{0.7ex plus .5ex minus 0ex}{\normalfont\normalsize\itshape}}%
\newlength{\CPheadmatchindent}% 
\def\@seccntformat#1{\hbox to\CPheadmatchindent{\csname the#1dis\endcsname}\hskip 0.1em \relax}
\IEEEilabelindentA \parindent
\IEEEilabelindent \IEEEilabelindentA
\IEEEelabelindent \parindent
\IEEEdlabelindent \parindent
\IEEElabelindent \parindent
\makeatother

\begin{document}
%%%%%%%%%%%%%%%%%%%%%%%%%%%%%%%%%%%%%%%%%%%%%%%%%%%%%%%%%%%%%%%%%%%%%%%%%%%%%
% We use \raggedbottom to avoid latex adding vertical space around headings.
% This gives a better idea to the author about how much white space remains
% as the page limit is approached.
\raggedbottom
%
%%%%%%%%%%%%%%%%%%%%%%%%%%%%%%%%%%%%%%%%%%%%%%%%%%%%%%%%%%%%%%%%%%%%%%%%%%%%%
% PAPER TITLE AND AUTHOR BLOCK
%
% The paper title can use linebreaks \\ within to get better formatting if desired.
%
\bstctlcite{IEEEexample:BSTcontrol}

\title{Experimental Evaluation of Moving Target Compensation in High Time-Bandwidth Noise Radar}
%
% Next we define the author names and affiliations.
% Author names are listed using \IEEEauthorblockN{} with comma separators between names.
% Affiliations are listed using \IEEEauthorblock{} with \\ separators between affiliations.
% Symbols marking author-affiliation relations are output using \EuMWauthorrefmark{}.
% At the end of the affiliation list is the list of author emails.
% See below for examples of each of these.
%
\author{
\IEEEauthorblockN{Martin Ankel$^{*1,2}$, Robert S. Jonsson$^{1,2}$, Mats Tholén$^{3,4}$,  Tomas Bryllert$^{1,2}$, Lars M.H. Ulander$^5$, Per Delsing$^1$} 
\\
\IEEEauthorblockA{
$^1$Department of Microtechnology and Nanoscience, Chalmers University of Technology, Sweden \\
$^2$Research and Concepts, Surveillance, 	Saab, Sweden  \\
$^3$Nanostructure Physics, KTH Royal Institute of Technology, Sweden\\
$^4$Intermodulation Products AB, Sweden
\\
$^5$Department of Space, Earth and Environment, Geoscience and Remote Sensing, Chalmers University of Technology, Sweden \\
 $^*$ankel@chalmers.se}
}

\IEEEoverridecommandlockouts

%
% Next we make the title/author block using the information defined above.
\maketitle
%%%%%%%%%%%%%%%%%%%%%%%%%%%%%%%%%%%%%%%%%%%%%%%%%%%%%%%%%%%%%%%%%%%%%%%%%%%%%
% ABSTRACT paragraph.
%
% As a general rule, do not put math, special symbols or citations
% in the abstract paragraph.
%
\begin{abstract} 
In this article, the effect a moving target has on the signal-to-interference-plus-noise-ratio (SINR) for high time-bandwidth noise radars is investigated. To compensate for cell migration we apply a computationally efficient stretch processing algorithm that is tailored for batched processing and suitable for implementation onto a real-time radar processor. The performance of the algorithm is studied using experimental data. In the experiment, pseudorandom noise, with a bandwidth of $100$~MHz, is generated and transmitted in real-time. An unmanned aerial vehicle (UAV), flown at a speed of $11.5$~m/s, is acting as a target. For an integration time of $1$~s, the algorithm is shown to yield an increase in SINR of roughly $13$~dB, compared to no compensation. It is also shown that coherent integration times of $2.5$~s can be achieved.

\end{abstract}


\IEEEpeerreviewmaketitle

% Effects of Movement for High Time-Bandwidths in Batched - Bra
% Coherent Integration With Range Migration
%Using Keystone Formatting - SMART kolla
%Pulse Compression Range-Doppler Radar - Relevant
% https://ieeexplore.ieee.org/stamp/stamp.jsp?tp=&arnumber=1180069 värd att läsa. Frequency dependent phase filter (det vi gör)
%  MIMO Radar with Widely Separated Antennas. - Kika på någon gång

%https://ieeexplore.ieee.org/stamp/stamp.jsp?tp=&arnumber=7925375 - complex stretch processing för gausiskt brus

\section{Introduction}
The military surveillance radar is a crucial piece of equipment in the modern battlefield, but it is expensive and vulnerable and, therefore, it must be heavily protected, especially as modern lightweight electronic warfare senors make precise and fast localisation of radar systems possible. Hence, methods and techniques of protecting the radar system is of high interest. One way to protect the radar system is to reduce the risk of the transmitted signal being intercepted in the first place, e.g., low probability of intercept (LPI) radar~\cite{LPIRadarStrategies, pace2009detecting}.  

To achieve LPI it is desirable to transmit a waveform with low spectral power density, i.e., the radar should transmit continuously with high instantaneous bandwidth and preferably integrate for an extended period of time. A type of radar that generally operates under these conditions is the noise radar~\cite{Horton, GeneralizedAmbiguityfunction, SuneAmbiguity, PulseShaping}. Introducing randomness to the waveform  makes it robust against intelligent jamming and, even if detected, the information revealed  is limited. For example, the radar mode of operation is difficult to deduce, since  the pulse repetition frequency is not revealed as it would be in most other types of radar systems. Additionally, noise radars are  unambiguous in range and Doppler~\cite{GeneralizedAmbiguityfunction, SuneAmbiguity}.

The combination of high bandwidth and long integration time will lead to significant  cell migration of a moving target during the coherent processing interval (CPI) and a reduction in SINR. As an example, with a bandwidth of $100$~MHz and a target moving at $10$~m/s the maximum coherent integration time is $150$~ms. To integrate for longer times, the movement of the target has to be accounted for. This can be done in the correlation processing by either compressing or expanding the reference signal, a method  referred to as stretch processing~\cite{Stretch:TimeTranformation, FastPulseDopplerRadarProcessing, StretchProcessingLongIntTime,  xu2011radon, Shan2016, EffectsofMovementforHighTimeBandwidths}. It should be mentioned that, while the velocity and acceleration of targets can be adequately accounted for, there may be  other sources of coherence loss that limit the coherent integration time.

The article is organised as follows. Section~\ref{sec:Theory} discusses the Doppler tolerance of noise radars and details a stretch processing algorithm tailored for batched processing of the noise signal. The experimental results are presented and discussed in Section~\ref{sec:Experiments}. Finally, the conclusions are given in Section~\ref{sec:Conclusions}.

\section{Theory Velocity Compensation}
\label{sec:Theory}
Noise radars generate and transmit a random or pseudorandom noise signal. Detection is performed by the concept of matched filter, which is implemented by performing a convolution between the received signal, $y$, and the reference signal, $x$, as
 \begin{equation}
    R[k] = \sum_n y_{n}x^*_{n-k},
     \label{eq:Cross}
\end{equation}
where the subscript denotes discrete samples of the respective signals. A correlation detector  contains the function $|R[k]|^2$ which will exhibit peaks at indices corresponding to strong correlations between the signal and the reference, indicating a reflection corresponding to that delay.
The magnitude of the peak depends on the similarity of $x$ and
$y$, meaning any distortion of $y$ will result in a correlation loss
and reduce the radar's detection performance. One such distortion is due to the movement of a target.

We assume that a target at a distance $R_0$ is moving with constant radial velocity $v_s$. The received signal at time $t$ is proportional to the reference delayed by $ 2(R_0-v_st)/c $, where $c$ is the speed of light. If the distance covered by the target during a CPI is insignificant in relation to the range resolution, only the Doppler shift has to be accounted for, and the term $2v_st/c$ can be ignored. However, if the covered distance matters, stretch processing has to be considered. These two cases will be treated separately and referred to as \textit{Doppler Compensation} and \textit{Stretch Compensation}, respectively.

\subsection{Doppler Compensation}
\label{sec:DopplerCompensation}
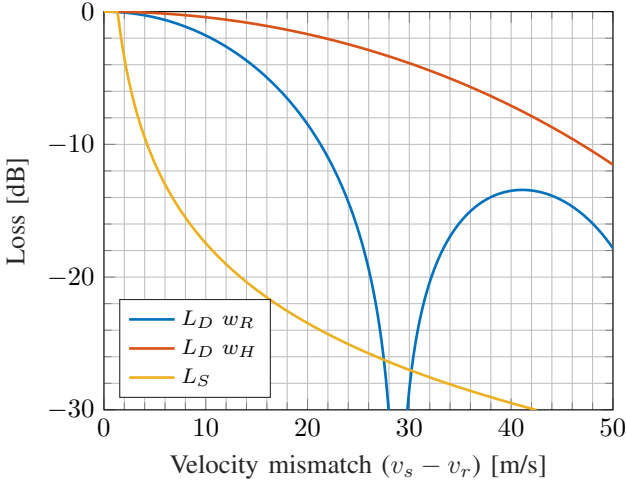
\begin{figure}
     \centering
     \begin{subfigure}[b]{0.45\textwidth}
         \centering
         % This file was created by matlab2tikz.
%
%The latest updates can be retrieved from
%  http://www.mathworks.com/matlabcentral/fileexchange/22022-matlab2tikz-matlab2tikz
%where you can also make suggestions and rate matlab2tikz.
%
\definecolor{mycolor1}{rgb}{0.00000,0.44700,0.74100}%
\definecolor{mycolor2}{rgb}{0.85000,0.32500,0.09800}%
\definecolor{mycolor3}{rgb}{0.92900,0.69400,0.12500}%
\begin{tikzpicture}

\begin{axis}[%
width=0.808\linewidth,
height=0.637\linewidth,
at={(0\linewidth,0\linewidth)},
scale only axis,
unbounded coords=jump,
xmin=0,
xmax=50,
xlabel style={font=\color{white!15!black}},
xlabel={Velocity mismatch ($v_s - v_r$) [m/s]},
ymin=-30,
ymax=0,
ylabel style={font=\color{white!15!black}},
ylabel={Loss [dB]},
axis background/.style={fill=white},
minor tick num=4,
xmajorgrids,
xminorgrids,
ymajorgrids,
yminorgrids,
legend style={at={(0.03,0.03)}, anchor=south west, legend cell align=left, align=left, draw=white!15!black, font = \footnotesize}
]
\addplot [color=mycolor1, line width=1.0pt]
  table[row sep=crcr]{%
0	0\\
0.799999999999997	-0.0110135291811275\\
1.6	-0.044087620987753\\
2.4	-0.0993232744806605\\
3.2	-0.176890457981671\\
4	-0.277030619558595\\
4.8	-0.400060316250503\\
5.6	-0.546376084422583\\
6.4	-0.716460720591684\\
7.2	-0.910891199471777\\
8	-1.13034852804459\\
8.8	-1.37562992685181\\
9.6	-1.6476638503208\\
10.3	-1.90848249007112\\
11	-2.19143618899145\\
11.7	-2.49747015328689\\
12.4	-2.82766256122148\\
13.1	-3.18324594454209\\
13.8	-3.56563351557561\\
14.5	-3.97645184058428\\
15.2	-4.41758175174076\\
15.8	-4.82146671910041\\
16.4	-5.25081347905964\\
17	-5.7074526004219\\
17.6	-6.19350261244936\\
18.2	-6.71143219364002\\
18.8	-7.2641404062712\\
19.3	-7.75375622873545\\
19.8	-8.27223475635451\\
20.3	-8.82231116063267\\
20.8	-9.40721338692114\\
21.3	-10.0307880596076\\
21.8	-10.6976695613758\\
22.2	-11.2661129340091\\
22.6	-11.8693592723731\\
23	-12.5115933638071\\
23.4	-13.1978865520288\\
23.8	-13.9344670124018\\
24.2	-14.7291017216358\\
24.5	-15.3690073342285\\
24.8	-16.0522419671033\\
25.1	-16.7851877285821\\
25.4	-17.5758217647473\\
25.7	-18.4342972769771\\
26	-19.3738158911636\\
26.3	-20.411986412171\\
26.5	-21.170499934978\\
26.7	-21.9929396833202\\
26.9	-22.8916141754526\\
27.1	-23.8828279419173\\
27.3	-24.9888234543665\\
27.5	-26.2410672463108\\
27.7	-27.6861779179583\\
27.9	-29.3975388360574\\
28	-30.3889908006127\\
nan	nan\\
29.8	-30.0084856314923\\
30	-28.4144977484505\\
30.2	-27.0883505805835\\
30.4	-25.9560865053641\\
30.6	-24.9707802735764\\
30.8	-24.1007894546493\\
31	-23.3237657721419\\
31.2	-22.6233392401889\\
31.4	-21.9871604693026\\
31.6	-21.4056839380045\\
31.9	-20.6199802716021\\
32.2	-19.9212262405838\\
32.5	-19.2949348586844\\
32.8	-18.730067246365\\
33.1	-18.2180211582575\\
33.4	-17.751965318936\\
33.7	-17.326386746319\\
34	-16.9367741933098\\
34.3	-16.5793913761927\\
34.6	-16.2511110733613\\
34.9	-15.949291491943\\
35.3	-15.5841847899429\\
35.7	-15.2576808753899\\
36.1	-14.9660922748261\\
36.5	-14.706351607811\\
36.9	-14.475886911527\\
37.3	-14.2725274059188\\
37.7	-14.0944312605957\\
38.1	-13.9400295425988\\
38.5	-13.8079822545012\\
38.9	-13.6971435401773\\
39.3	-13.6065339389204\\
39.7	-13.5353181309264\\
40.2	-13.4725014433406\\
40.7	-13.4377451332141\\
41.2	-13.4301453406245\\
41.7	-13.4490107131309\\
42.2	-13.4938429856112\\
42.7	-13.5643232465012\\
43.2	-13.660303108584\\
43.7	-13.781800295538\\
44.2	-13.9289984015464\\
44.7	-14.1022508044202\\
45.2	-14.3020889335725\\
45.7	-14.5292353330999\\
46.2	-14.7846222399402\\
46.7	-15.0694167462232\\
47.2	-15.3850540725531\\
47.7	-15.7332811013498\\
48.1	-16.0367346599084\\
48.5	-16.3636739280869\\
48.9	-16.7155936722198\\
49.3	-17.0942394288541\\
49.7	-17.5016567255811\\
50	-17.8275339889241\\
};
\addlegendentry{$L_D$ $w_R$}

\addplot [color=mycolor2, line width=1.0pt]
  table[row sep=crcr]{%
0	0\\
1.7	-0.0120896337992491\\
3.4	-0.0483721126333947\\
5.1	-0.108888259814222\\
6.7	-0.188042804473049\\
8.3	-0.28880376265937\\
9.9	-0.411273178195408\\
11.5	-0.55557615691081\\
13.1	-0.72186164557499\\
14.7	-0.910303376587983\\
16.3	-1.1211009938493\\
17.9	-1.35448137871541\\
19.5	-1.61070019904626\\
21.1	-1.89004370918588\\
22.7	-2.19283083451806\\
24.3	-2.5194155812323\\
25.9	-2.87018982044483\\
27.5	-3.24558650624998\\
29	-3.62030727838009\\
30.5	-4.0175211012148\\
32	-4.43770228674859\\
33.5	-4.88137088219529\\
35	-5.34909751404277\\
36.5	-5.84150902402215\\
38	-6.35929505421489\\
39.4	-6.86612493600725\\
40.8	-7.39640067032491\\
42.2	-7.95087065809462\\
43.6	-8.53036030078594\\
45	-9.13578264121468\\
46.3	-9.72206491300527\\
47.6	-10.3324802667555\\
48.9	-10.968027295312\\
50	-11.5262478745149\\
};
\addlegendentry{$L_D$  $w_H$}

\addplot [color=mycolor3, line width=1.0pt]
  table[row sep=crcr]{%
0	-0\\
1.3	-0\\
1.4	-0.369835662531699\\
1.6	-1.52967460208544\\
1.8	-2.55272505103306\\
2	-3.46787486224656\\
2.2	-4.29572856541107\\
2.4	-5.05149978319906\\
2.7	-6.07455023214668\\
3	-6.98970004336019\\
3.3	-7.81755374652469\\
3.6	-8.57332496431269\\
3.9	-9.26856708949693\\
4.2	-9.91226075692495\\
4.5	-10.5115252244738\\
4.9	-11.2511965495372\\
5.3	-11.9327923409827\\
5.7	-12.5647720624168\\
6.1	-13.1538716491823\\
6.5	-13.705542081824\\
6.9	-14.224256763712\\
7.4	-14.8319093435865\\
7.9	-15.3998167747758\\
8.4	-15.9328606702046\\
8.9	-16.4350750818652\\
9.5	-17.0017470547439\\
10.1	-17.5337024246198\\
10.7	-18.0349505026711\\
11.3	-18.5088438186353\\
12	-19.0308998699194\\
12.7	-19.5233493680861\\
13.4	-19.9893709162631\\
14.2	-20.4930418366281\\
15	-20.9691001300806\\
15.8	-21.4204166880554\\
16.7	-21.9016043719186\\
17.6	-22.3575283052499\\
18.6	-22.8375338333253\\
19.6	-23.2923963760965\\
20.7	-23.7666818581053\\
21.8	-24.216404821059\\
23	-24.6818316693188\\
24.2	-25.1235822685756\\
25.5	-25.578078557646\\
26.9	-26.0423205490151\\
28.3	-26.4830036594527\\
29.8	-26.931600230492\\
31.4	-27.3858679104312\\
33.1	-27.8438348244813\\
34.9	-28.3037834881505\\
36.8	-28.7642313224373\\
38.8	-29.2239094608511\\
40.9	-29.6817411091138\\
42.5	-30.0150535499732\\
};
\addlegendentry{$L_S$}

\end{axis}

\begin{axis}[%
width=1.043\linewidth,
height=0.782\linewidth,
at={(-0.136\linewidth,-0.086\linewidth)},
scale only axis,
xmin=0,
xmax=1,
ymin=0,
ymax=1,
axis line style={draw=none},
ticks=none,
axis x line*=bottom,
axis y line*=left
]
\end{axis}
\end{tikzpicture}%
     \end{subfigure}
     \hfill
        \caption{Theoretical Doppler loss ($L_D$), for a rectangular window ($w_R$) and a Hann window ($w_H$), and theoretical cell migration loss ($L_S$) as a function of velocity mismatch between the reference and signal. The losses are calculated for a carrier frequency of $1.3$~GHz, a batch length of $4$~ms, a bandwidth of $100$~MHz, a sampling rate of $125$~MHz and a CPI length of $1$~s.  }     
        \label{fig:FBandCorrTimes}
\end{figure}

Compensating for the  Doppler shift is performed by applying a  Doppler modulation to the reference signal as
\begin{equation}
   x'_n = x_n \cdot \mathrm{e}^{-2\pi {\rm i}(2v_rf_c/c)\cdot \frac{n}{f_s}},
   \label{eq:DopplerCompensation}
\end{equation}
where $v_r$ is the reference velocity, $f_s$ is the receiver's sampling rate and $f_c$ is the carrier frequency. If $v_r$ is close to the target velocity, the loss due to the movement of the target is  mitigated. In a real world application the target velocity is not known to the radar and must be guessed. Performing a calculation for every possible velocity would be extremely resource demanding. A more efficient approach is to compute the correlations on a grid of different velocities spaced by $\delta v$, where $\delta v$ should be selected such that the maximum Doppler loss does not exceed a specified threshold. 

The correlation loss due to a Doppler shift for noise radars has been investigated in~\cite{DopplerTolerance} and it was found that the loss primarily depends on the window function $w(t)$ as
\begin{equation}
    L_{D} =  \Big{|}\mathcal{F}(w(t))\Big{|}^{-2}.
    \label{eq:DopplerLoss}
\end{equation}
For example, for a rectangular window the loss is 
\begin{equation}
   L_{D} = \left|\frac{\sin[2\pi f_c (v_s-v_r)t_p/c]}{2\pi f_c (v_s-v_r)t_p/c} \right|^{-2}, 
\end{equation}
where $t_p$ is the batch length. To get an idea of the velocity spacing required we simulate the loss at L-band ($1.3$~GHz) in Fig.~\ref{fig:FBandCorrTimes}, for a rectangular window as well as a Hann window, with $t_p=4$~ms. In this case, if $3$~dB loss is acceptable, $\delta v$ would be roughly $26$~m/s for a rectangular window and $53$~m/s for a Hann window. It should be kept in mind  that a Hann window also decreases the signal to noise ratio with approximately $1.8$~dB. The $\delta v$ spacing basically scales linearly with batch length, meaning shorter batches of $0.4$~ms and $40$~\textmu s would have a required spacing of $260$~m/s and $2600$~m/s, respectively, for a rectangular window. %Of course, higher frequencies sets stricter requirements on $\delta v$.

\subsection{Stretch Compensation}
\label{sec:stretch}
Typically, in order to calculate the range-Doppler map for a noise waveform in an efficient manner, batched processing is preferred~\cite{ BatchProcessing2}, especially for real-time implementation on, e.g., a field programmable gate array (FPGA). Therefore, it is desirable to implement the stretch compensation in conjunction with the batched operation, while simultaneously keeping the number of extra operations to a minimum and avoiding the need to buffer data.

Here, we present an algorithm that efficiently performs an approximate stretch processing by applying a constant phase factor per batch~\cite{EffectsofMovementforHighTimeBandwidths, Shan2016}. Assume that the CPI consists of $N$ samples, labelled by index $n = 0,1,\ldots, N-1$, and that the Doppler shifted reference signal, $x'_n$ in \eqref{eq:DopplerCompensation}, is segmented into $P$ batches, each batch containing $M = N/P$ samples covering a time of $t_p$. Let each batch populate a row in a matrix, $\mathbf{x'}_{p,m}$, of size $P\times M$ and form the discrete Fourier transform of each respective row, as
\begin{equation}
   \mathbf{X}_{p,q} = \frac{1}{M}\sum_{m=0}^{M-1} \mathbf{x}'_{p,m}\mathrm{e}^{-2\pi\mathrm{i}  \frac{q \cdot  m}{M}}.
\end{equation}
If the distance the target moves during the time $t_p$ is negligible with respect to the range resolution, stretching over each individual batch is unnecessary. In this case it is sufficient to  perform time
translation only in slow time, i.e., between batches. This can be done in the frequency domain, utilising the Fourier transform property that $\mathcal{F}\left[f(n-a)\right] = \mathcal{F}\left[f(n)\right]\mathrm{e}^{- \frac{2\pi\mathrm{i}}{N}ka  }$.  Between batches, the reference is  shifted with the factor $a=2v_rf_st_p/c$ to compensate for target motion, as
\begin{equation}
   \mathbf{X}_{p,q,r}^{'} = \mathbf{X}_{p,q}\mathrm{e}^{-2\pi\mathrm{i} \frac{q\cdot p}{M} \frac{2v_rf_st_p}{c}} = \mathbf{X}_{p,q}\mathrm{e}^{-2\pi\mathrm{i} q\cdot p \frac{2v_r}{c}},
    \label{eq:StretchedREf}
\end{equation}
where $p = 0,1,\ldots,P-1$ is the batch index and where we have utilised the fact that $t_p = M/f_s$. 

The range-Doppler map is calculated by performing the cross-correlation in the frequency domain and then taking the Fourier transform of each column, i.e.,
\begin{equation}
    \mathbf{R}_{l,m,r} = \sum_{p=0}^{P-1}\Biggl( \sum_{q=0}^{M-1} \mathbf{Y}_{p,q}\left(\mathbf{X}'_{p,q,r}\right)^*\mathrm{e}^{2\pi \mathrm{i} \frac{q\cdot p}{M}}\Biggr)\mathrm{e}^{-2\pi \mathrm{i} \frac{l\cdot p}{P}}.
\end{equation}
This processing is efficiently implemented with the Fast Fourier Transform algorithm. There are only two additional operations required by the algorithm to perform the stretch compensation: calculation of one phase factor for each sample and one element-wise multiplication. In FPGA terms this would consist of a cosine block, a sine block and two product blocks (and some counters), making it easy to implement and relatively resource efficient. Such an implementation is also practical considering data does not have to be buffered, but can be continuously streamed. 

The spacing of the different stretch hypotheses $dv$ can be determined by requiring a maximum allowable stretch loss $L_S$,  given by~\cite{EffectsofMovementforHighTimeBandwidths}
\begin{equation}
    L_{S} = 4(v_s-v_r)^2T_{\rm  int}^2f_sB/c^2,
    \label{eq:StretchLoss}
\end{equation}
where $T_{\rm int}$ is the integration time. The migration loss as function of velocity mismatch can be seen in Fig.~\ref{fig:FBandCorrTimes}, for $f_s = 125$~MHz, $T_{\rm int} = 1$~s and $B = 100$~MHz. If $3$~dB loss is considered to be acceptable, the required spacing is $dv =1.9$~m/s. This is quite  a dense grid that would require considerable computational resources. Luckily, there are several ways the calculations can be factorised to reduce the computational load, see e.g.,~\cite{UlanderFactorizedBackProjection}. 

In summary, it is important to keep track of both the Doppler loss \eqref{eq:DopplerLoss} and the stretch loss \eqref{eq:StretchLoss} when designing the system and choosing $\delta v$ and $dv$.

\section{Experiments and Results}
\label{sec:Experiments}
In the experiments a pseudorandom sequence of $100$~MHz bandwidth is generated and transmitted in real-time, at the L-band ($1.3$~GHz), by a digital microwave platform named Vivace~\cite{Vivace}. 
Data is recorded with a software defined radio, capable of streaming data to disk drive storage at a rate of $250$~MS/s. The equipment used is detailed in~\cite{DemonstrationofCorrelationNoiseSuppression}, with the only difference being that the transmitter's and receiver's FPGAs have been upgraded to handle continuous transmission and reception of non-repeating signals. Acting as a target is a DJI Matrice 600 UAV, flown in the direction of the antenna boresight at a speed of $11.5$~m/s for $5$~s.%, during which  it covered a distance of $80$~m. %To minimise wind influence, emphasise was put on conducting the experiment on a windless day.

\begin{figure}
    \centering
    \begin{subfigure}[b]{0.45\textwidth}
            \centering
            \input{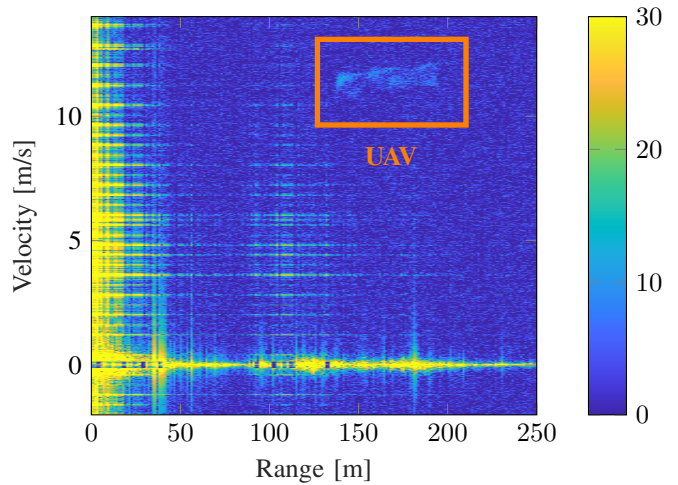}
    \end{subfigure}
    \caption{Range-Doppler map of the entire, cluttered filtered, $5$~s recorded sequence. The Doppler sidelobes are both due to spectral leakage from  strong clutter scatterers (direct signal, ground returns, power lines, etc.) as well as due to spurious frequencies
    originating from the digital to analog converter. The UAV signal is highlighted by an orange box. }
    \label{fig:IntegratedAllPulses}
\end{figure}

Since this scenario consists of a monostatic continuous wave noise radar, there is significant self interference present,  which is masking the target~\cite{MaskingeffectSune}. To be able to  detect the UAV we processed the data with the Sequential CLEAN algorithm~\cite{DemonstrationofCorrelationNoiseSuppression} to suppress strong clutter scatterers. When integrating the cleaned data over the entire recording with a batch time of $t_p=4$~ms, we get the results shown in Fig.~\ref{fig:IntegratedAllPulses}. The UAV is detectable with a maximum SINR of roughly $15.6$~dB, where the $0$~dB level is referenced to the average interference-plus-noise floor. As expected, there is significant broadening of the target signal in both range and Doppler.

\begin{figure*}
     \begin{subfigure}[b]{0.24\textwidth}
         \centering
         % This file was created by matlab2tikz.
%
%The latest updates can be retrieved from
%  http://www.mathworks.com/matlabcentral/fileexchange/22022-matlab2tikz-matlab2tikz
%where you can also make suggestions and rate matlab2tikz.
%
\begin{tikzpicture}

\begin{axis}[%
width=0.7\linewidth,
height=0.6\linewidth,
at={(0\linewidth,0\linewidth)},
scale only axis,
point meta min=0,
point meta max=30,
axis on top,
xmin=155,
xmax=175,
xlabel style={font=\color{white!15!black}},
xlabel={Range [m]},
y dir=reverse,
ymin=-12.5,
ymax=-10.5,
ylabel style={font=\color{white!15!black}},
ylabel={Velocity [m/s]},
xtick={160, 165, 170},
ytick={-12, -11.5, -11},
yticklabels = {12, 11.5, 11},
label style = {font = \footnotesize},
tick label style={font=\scriptsize},
axis background/.style={fill=white},
colormap={mymap}{[1pt] rgb(0pt)=(0.2422,0.1504,0.6603); rgb(1pt)=(0.2444,0.1534,0.6728); rgb(2pt)=(0.2464,0.1569,0.6847); rgb(3pt)=(0.2484,0.1607,0.6961); rgb(4pt)=(0.2503,0.1648,0.7071); rgb(5pt)=(0.2522,0.1689,0.7179); rgb(6pt)=(0.254,0.1732,0.7286); rgb(7pt)=(0.2558,0.1773,0.7393); rgb(8pt)=(0.2576,0.1814,0.7501); rgb(9pt)=(0.2594,0.1854,0.761); rgb(11pt)=(0.2628,0.1932,0.7828); rgb(12pt)=(0.2645,0.1972,0.7937); rgb(13pt)=(0.2661,0.2011,0.8043); rgb(14pt)=(0.2676,0.2052,0.8148); rgb(15pt)=(0.2691,0.2094,0.8249); rgb(16pt)=(0.2704,0.2138,0.8346); rgb(17pt)=(0.2717,0.2184,0.8439); rgb(18pt)=(0.2729,0.2231,0.8528); rgb(19pt)=(0.274,0.228,0.8612); rgb(20pt)=(0.2749,0.233,0.8692); rgb(21pt)=(0.2758,0.2382,0.8767); rgb(22pt)=(0.2766,0.2435,0.884); rgb(23pt)=(0.2774,0.2489,0.8908); rgb(24pt)=(0.2781,0.2543,0.8973); rgb(25pt)=(0.2788,0.2598,0.9035); rgb(26pt)=(0.2794,0.2653,0.9094); rgb(27pt)=(0.2798,0.2708,0.915); rgb(28pt)=(0.2802,0.2764,0.9204); rgb(29pt)=(0.2806,0.2819,0.9255); rgb(30pt)=(0.2809,0.2875,0.9305); rgb(31pt)=(0.2811,0.293,0.9352); rgb(32pt)=(0.2813,0.2985,0.9397); rgb(33pt)=(0.2814,0.304,0.9441); rgb(34pt)=(0.2814,0.3095,0.9483); rgb(35pt)=(0.2813,0.315,0.9524); rgb(36pt)=(0.2811,0.3204,0.9563); rgb(37pt)=(0.2809,0.3259,0.96); rgb(38pt)=(0.2807,0.3313,0.9636); rgb(39pt)=(0.2803,0.3367,0.967); rgb(40pt)=(0.2798,0.3421,0.9702); rgb(41pt)=(0.2791,0.3475,0.9733); rgb(42pt)=(0.2784,0.3529,0.9763); rgb(43pt)=(0.2776,0.3583,0.9791); rgb(44pt)=(0.2766,0.3638,0.9817); rgb(45pt)=(0.2754,0.3693,0.984); rgb(46pt)=(0.2741,0.3748,0.9862); rgb(47pt)=(0.2726,0.3804,0.9881); rgb(48pt)=(0.271,0.386,0.9898); rgb(49pt)=(0.2691,0.3916,0.9912); rgb(50pt)=(0.267,0.3973,0.9924); rgb(51pt)=(0.2647,0.403,0.9935); rgb(52pt)=(0.2621,0.4088,0.9946); rgb(53pt)=(0.2591,0.4145,0.9955); rgb(54pt)=(0.2556,0.4203,0.9965); rgb(55pt)=(0.2517,0.4261,0.9974); rgb(56pt)=(0.2473,0.4319,0.9983); rgb(57pt)=(0.2424,0.4378,0.9991); rgb(58pt)=(0.2369,0.4437,0.9996); rgb(59pt)=(0.2311,0.4497,0.9995); rgb(60pt)=(0.225,0.4559,0.9985); rgb(61pt)=(0.2189,0.462,0.9968); rgb(62pt)=(0.2128,0.4682,0.9948); rgb(63pt)=(0.2066,0.4743,0.9926); rgb(64pt)=(0.2006,0.4803,0.9906); rgb(65pt)=(0.195,0.4861,0.9887); rgb(66pt)=(0.1903,0.4919,0.9867); rgb(67pt)=(0.1869,0.4975,0.9844); rgb(68pt)=(0.1847,0.503,0.9819); rgb(69pt)=(0.1831,0.5084,0.9793); rgb(70pt)=(0.1818,0.5138,0.9766); rgb(71pt)=(0.1806,0.5191,0.9738); rgb(72pt)=(0.1795,0.5244,0.9709); rgb(73pt)=(0.1785,0.5296,0.9677); rgb(74pt)=(0.1778,0.5349,0.9641); rgb(75pt)=(0.1773,0.5401,0.9602); rgb(76pt)=(0.1768,0.5452,0.956); rgb(77pt)=(0.1764,0.5504,0.9516); rgb(78pt)=(0.1755,0.5554,0.9473); rgb(79pt)=(0.174,0.5605,0.9432); rgb(80pt)=(0.1716,0.5655,0.9393); rgb(81pt)=(0.1686,0.5705,0.9357); rgb(82pt)=(0.1649,0.5755,0.9323); rgb(83pt)=(0.161,0.5805,0.9289); rgb(84pt)=(0.1573,0.5854,0.9254); rgb(85pt)=(0.154,0.5902,0.9218); rgb(86pt)=(0.1513,0.595,0.9182); rgb(87pt)=(0.1492,0.5997,0.9147); rgb(88pt)=(0.1475,0.6043,0.9113); rgb(89pt)=(0.1461,0.6089,0.908); rgb(90pt)=(0.1446,0.6135,0.905); rgb(91pt)=(0.1429,0.618,0.9022); rgb(92pt)=(0.1408,0.6226,0.8998); rgb(93pt)=(0.1383,0.6272,0.8975); rgb(94pt)=(0.1354,0.6317,0.8953); rgb(95pt)=(0.1321,0.6363,0.8932); rgb(96pt)=(0.1288,0.6408,0.891); rgb(97pt)=(0.1253,0.6453,0.8887); rgb(98pt)=(0.1219,0.6497,0.8862); rgb(99pt)=(0.1185,0.6541,0.8834); rgb(100pt)=(0.1152,0.6584,0.8804); rgb(101pt)=(0.1119,0.6627,0.877); rgb(102pt)=(0.1085,0.6669,0.8734); rgb(103pt)=(0.1048,0.671,0.8695); rgb(104pt)=(0.1009,0.675,0.8653); rgb(105pt)=(0.0964,0.6789,0.8609); rgb(106pt)=(0.0914,0.6828,0.8562); rgb(107pt)=(0.0855,0.6865,0.8513); rgb(108pt)=(0.0789,0.6902,0.8462); rgb(109pt)=(0.0713,0.6938,0.8409); rgb(110pt)=(0.0628,0.6972,0.8355); rgb(111pt)=(0.0535,0.7006,0.8299); rgb(112pt)=(0.0433,0.7039,0.8242); rgb(113pt)=(0.0328,0.7071,0.8183); rgb(114pt)=(0.0234,0.7103,0.8124); rgb(115pt)=(0.0155,0.7133,0.8064); rgb(116pt)=(0.0091,0.7163,0.8003); rgb(117pt)=(0.0046,0.7192,0.7941); rgb(118pt)=(0.0019,0.722,0.7878); rgb(119pt)=(0.0009,0.7248,0.7815); rgb(120pt)=(0.0018,0.7275,0.7752); rgb(121pt)=(0.0046,0.7301,0.7688); rgb(122pt)=(0.0094,0.7327,0.7623); rgb(123pt)=(0.0162,0.7352,0.7558); rgb(124pt)=(0.0253,0.7376,0.7492); rgb(125pt)=(0.0369,0.74,0.7426); rgb(126pt)=(0.0504,0.7423,0.7359); rgb(127pt)=(0.0638,0.7446,0.7292); rgb(128pt)=(0.077,0.7468,0.7224); rgb(129pt)=(0.0899,0.7489,0.7156); rgb(130pt)=(0.1023,0.751,0.7088); rgb(131pt)=(0.1141,0.7531,0.7019); rgb(132pt)=(0.1252,0.7552,0.695); rgb(133pt)=(0.1354,0.7572,0.6881); rgb(134pt)=(0.1448,0.7593,0.6812); rgb(135pt)=(0.1532,0.7614,0.6741); rgb(136pt)=(0.1609,0.7635,0.6671); rgb(137pt)=(0.1678,0.7656,0.6599); rgb(138pt)=(0.1741,0.7678,0.6527); rgb(139pt)=(0.1799,0.7699,0.6454); rgb(140pt)=(0.1853,0.7721,0.6379); rgb(141pt)=(0.1905,0.7743,0.6303); rgb(142pt)=(0.1954,0.7765,0.6225); rgb(143pt)=(0.2003,0.7787,0.6146); rgb(144pt)=(0.2061,0.7808,0.6065); rgb(145pt)=(0.2118,0.7828,0.5983); rgb(146pt)=(0.2178,0.7849,0.5899); rgb(147pt)=(0.2244,0.7869,0.5813); rgb(148pt)=(0.2318,0.7887,0.5725); rgb(149pt)=(0.2401,0.7905,0.5636); rgb(150pt)=(0.2491,0.7922,0.5546); rgb(151pt)=(0.2589,0.7937,0.5454); rgb(152pt)=(0.2695,0.7951,0.536); rgb(153pt)=(0.2809,0.7964,0.5266); rgb(154pt)=(0.2929,0.7975,0.517); rgb(155pt)=(0.3052,0.7985,0.5074); rgb(156pt)=(0.3176,0.7994,0.4975); rgb(157pt)=(0.3301,0.8002,0.4876); rgb(158pt)=(0.3424,0.8009,0.4774); rgb(159pt)=(0.3548,0.8016,0.4669); rgb(160pt)=(0.3671,0.8021,0.4563); rgb(161pt)=(0.3795,0.8026,0.4454); rgb(162pt)=(0.3921,0.8029,0.4344); rgb(163pt)=(0.405,0.8031,0.4233); rgb(164pt)=(0.4184,0.803,0.4122); rgb(165pt)=(0.4322,0.8028,0.4013); rgb(166pt)=(0.4463,0.8024,0.3904); rgb(167pt)=(0.4608,0.8018,0.3797); rgb(168pt)=(0.4753,0.8011,0.3691); rgb(169pt)=(0.4899,0.8002,0.3586); rgb(170pt)=(0.5044,0.7993,0.348); rgb(171pt)=(0.5187,0.7982,0.3374); rgb(172pt)=(0.5329,0.797,0.3267); rgb(173pt)=(0.547,0.7957,0.3159); rgb(175pt)=(0.5748,0.7929,0.2941); rgb(176pt)=(0.5886,0.7913,0.2833); rgb(177pt)=(0.6024,0.7896,0.2726); rgb(178pt)=(0.6161,0.7878,0.2622); rgb(179pt)=(0.6297,0.7859,0.2521); rgb(180pt)=(0.6433,0.7839,0.2423); rgb(181pt)=(0.6567,0.7818,0.2329); rgb(182pt)=(0.6701,0.7796,0.2239); rgb(183pt)=(0.6833,0.7773,0.2155); rgb(184pt)=(0.6963,0.775,0.2075); rgb(185pt)=(0.7091,0.7727,0.1998); rgb(186pt)=(0.7218,0.7703,0.1924); rgb(187pt)=(0.7344,0.7679,0.1852); rgb(188pt)=(0.7468,0.7654,0.1782); rgb(189pt)=(0.759,0.7629,0.1717); rgb(190pt)=(0.771,0.7604,0.1658); rgb(191pt)=(0.7829,0.7579,0.1608); rgb(192pt)=(0.7945,0.7554,0.157); rgb(193pt)=(0.806,0.7529,0.1546); rgb(194pt)=(0.8172,0.7505,0.1535); rgb(195pt)=(0.8281,0.7481,0.1536); rgb(196pt)=(0.8389,0.7457,0.1546); rgb(197pt)=(0.8495,0.7435,0.1564); rgb(198pt)=(0.86,0.7413,0.1587); rgb(199pt)=(0.8703,0.7392,0.1615); rgb(200pt)=(0.8804,0.7372,0.165); rgb(201pt)=(0.8903,0.7353,0.1695); rgb(202pt)=(0.9,0.7336,0.1749); rgb(203pt)=(0.9093,0.7321,0.1815); rgb(204pt)=(0.9184,0.7308,0.189); rgb(205pt)=(0.9272,0.7298,0.1973); rgb(206pt)=(0.9357,0.729,0.2061); rgb(207pt)=(0.944,0.7285,0.2151); rgb(208pt)=(0.9523,0.7284,0.2237); rgb(209pt)=(0.9606,0.7285,0.2312); rgb(210pt)=(0.9689,0.7292,0.2373); rgb(211pt)=(0.977,0.7304,0.2418); rgb(212pt)=(0.9842,0.733,0.2446); rgb(213pt)=(0.99,0.7365,0.2429); rgb(214pt)=(0.9946,0.7407,0.2394); rgb(215pt)=(0.9966,0.7458,0.2351); rgb(216pt)=(0.9971,0.7513,0.2309); rgb(217pt)=(0.9972,0.7569,0.2267); rgb(218pt)=(0.9971,0.7626,0.2224); rgb(219pt)=(0.9969,0.7683,0.2181); rgb(220pt)=(0.9966,0.774,0.2138); rgb(221pt)=(0.9962,0.7798,0.2095); rgb(222pt)=(0.9957,0.7856,0.2053); rgb(223pt)=(0.9949,0.7915,0.2012); rgb(224pt)=(0.9938,0.7974,0.1974); rgb(225pt)=(0.9923,0.8034,0.1939); rgb(226pt)=(0.9906,0.8095,0.1906); rgb(227pt)=(0.9885,0.8156,0.1875); rgb(228pt)=(0.9861,0.8218,0.1846); rgb(229pt)=(0.9835,0.828,0.1817); rgb(230pt)=(0.9807,0.8342,0.1787); rgb(231pt)=(0.9778,0.8404,0.1757); rgb(232pt)=(0.9748,0.8467,0.1726); rgb(233pt)=(0.972,0.8529,0.1695); rgb(234pt)=(0.9694,0.8591,0.1665); rgb(235pt)=(0.9671,0.8654,0.1636); rgb(236pt)=(0.9651,0.8716,0.1608); rgb(237pt)=(0.9634,0.8778,0.1582); rgb(238pt)=(0.9619,0.884,0.1557); rgb(239pt)=(0.9608,0.8902,0.1532); rgb(240pt)=(0.9601,0.8963,0.1507); rgb(241pt)=(0.9596,0.9023,0.148); rgb(242pt)=(0.9595,0.9084,0.145); rgb(243pt)=(0.9597,0.9143,0.1418); rgb(244pt)=(0.9601,0.9203,0.1382); rgb(245pt)=(0.9608,0.9262,0.1344); rgb(246pt)=(0.9618,0.932,0.1304); rgb(247pt)=(0.9629,0.9379,0.1261); rgb(248pt)=(0.9642,0.9437,0.1216); rgb(249pt)=(0.9657,0.9494,0.1168); rgb(250pt)=(0.9674,0.9552,0.1116); rgb(251pt)=(0.9692,0.9609,0.1061); rgb(252pt)=(0.9711,0.9667,0.1001); rgb(253pt)=(0.973,0.9724,0.0938); rgb(254pt)=(0.9749,0.9782,0.0872); rgb(255pt)=(0.9769,0.9839,0.0805)}
]
\addplot [forget plot] graphics [xmin=-101.4, xmax=400.2, ymin=-14.4807692307692, ymax=14.3653846153846] {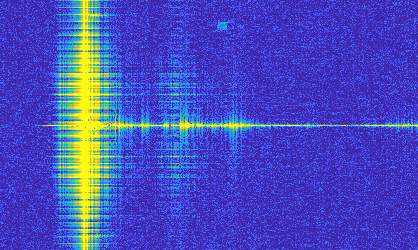};
\end{axis}

\begin{axis}[%
width=1.0\linewidth,
height=0.75\linewidth,
at={(0,0)},
scale only axis,
point meta min=0,
point meta max=1,
xmin=0,
xmax=1,
ymin=0,
ymax=1,
axis line style={draw=none},
ticks=none,
axis x line*=bottom,
axis y line*=left
]
\end{axis}
\end{tikzpicture}%
         \caption{}
         \label{fig:y equals x}
     \end{subfigure}
     \begin{subfigure}[b]{0.24\textwidth}
         \centering
         % This file was created by matlab2tikz.
%
%The latest updates can be retrieved from
%  http://www.mathworks.com/matlabcentral/fileexchange/22022-matlab2tikz-matlab2tikz
%where you can also make suggestions and rate matlab2tikz.
%
\begin{tikzpicture}

\begin{axis}[%
width=0.7\linewidth,
height=0.6\linewidth,
at={(0\linewidth,0\linewidth)},
scale only axis,
point meta min=0,
point meta max=30,
axis on top,
xmin=155,
xmax=175,
xlabel style={font=\color{white!15!black}},
xlabel={Range [m]},
y dir=reverse,
ymin=-1,
ymax=1,
xtick={160, 165, 170},
ytick={-0.5, 0, 0.5},
yticklabels = {12, 11.5, 11},
label style = {font = \footnotesize},
tick label style={font=\scriptsize},
axis background/.style={fill=white}
]
\addplot [forget plot] graphics [xmin=-101.4, xmax=400.2, ymin=-14.4807692307692, ymax=14.3653846153846] {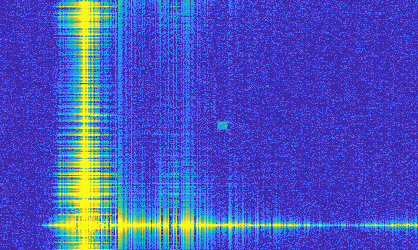};
\end{axis}

\begin{axis}[%
width=1.0\linewidth,
height=0.75\linewidth,
at={(-0.136\linewidth,-0.086\linewidth)},
scale only axis,
xmin=0,
xmax=1,
ymin=0,
ymax=1,
axis line style={draw=none},
ticks=none,
axis x line*=bottom,
axis y line*=left
]
\end{axis}
\end{tikzpicture}%
         \caption{}
         \label{fig:three sin x}
     \end{subfigure} 
     \hspace{-5mm}
     \begin{subfigure}[b]{0.24\textwidth}
         \centering
         % This file was created by matlab2tikz.
%
%The latest updates can be retrieved from
%  http://www.mathworks.com/matlabcentral/fileexchange/22022-matlab2tikz-matlab2tikz
%where you can also make suggestions and rate matlab2tikz.
%
\begin{tikzpicture}

\begin{axis}[%
width=0.7\linewidth,
height=0.6\linewidth,
at={(0\linewidth,0\linewidth)},
scale only axis,
point meta min=0,
point meta max=30,
axis on top,
xmin=155,
xmax=175,
xlabel style={font=\color{white!15!black}},
xlabel={Range [m]},
y dir=reverse,
ymin=-12.5,
ymax=-10.5,
ytick={\empty},
xtick = {160, 165, 170},
ytick = {-12, -11.5, -11},
yticklabels = {12, 11.5, 11},
label style = {font = \footnotesize},
tick label style={font=\scriptsize},
axis background/.style={fill=white}
]
\addplot [forget plot] graphics [xmin=-501, xmax=499.8, ymin=-14.4807692307692, ymax=14.3653846153846] {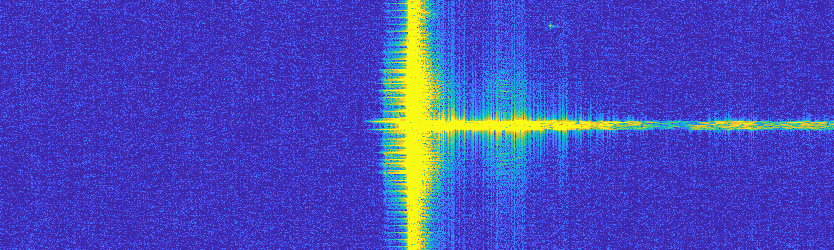};
\end{axis}

\begin{axis}[%
width=1.0\linewidth,
height=0.75\linewidth,
at={(-0.136\linewidth,-0.086\linewidth)},
scale only axis,
xmin=0,
xmax=1,
ymin=0,
ymax=1,
axis line style={draw=none},
ticks=none,
axis x line*=bottom,
axis y line*=left
]
\end{axis}
\end{tikzpicture}%
         \caption{}
         \label{fig:five over x}
     \end{subfigure}
    \hspace{-5mm}
      \begin{subfigure}[b]{0.24\textwidth}
         \centering
         % This file was created by matlab2tikz.
%
%The latest updates can be retrieved from
%  http://www.mathworks.com/matlabcentral/fileexchange/22022-matlab2tikz-matlab2tikz
%where you can also make suggestions and rate matlab2tikz.
%
\begin{tikzpicture}

\begin{axis}[%
width=0.7\linewidth,
height=0.6\linewidth,
at={(0\linewidth,0\linewidth)},
scale only axis,
point meta min=0,
point meta max=30,
axis on top,
xmin=155,
xmax=175,
xlabel style={font=\color{white!15!black}},
xlabel={Range [m]},
y dir=reverse,
ymin=-1,
ymax=1,
ytick={\empty},
xtick={160, 165, 170},
ytick={-0.5, 0, 0.5},
yticklabels = {12, 11.5, 11},
label style = {font = \footnotesize},
tick label style={font=\scriptsize},
axis background/.style={fill=white},
colormap={mymap}{[1pt] rgb(0pt)=(0.2422,0.1504,0.6603); rgb(1pt)=(0.2444,0.1534,0.6728); rgb(2pt)=(0.2464,0.1569,0.6847); rgb(3pt)=(0.2484,0.1607,0.6961); rgb(4pt)=(0.2503,0.1648,0.7071); rgb(5pt)=(0.2522,0.1689,0.7179); rgb(6pt)=(0.254,0.1732,0.7286); rgb(7pt)=(0.2558,0.1773,0.7393); rgb(8pt)=(0.2576,0.1814,0.7501); rgb(9pt)=(0.2594,0.1854,0.761); rgb(11pt)=(0.2628,0.1932,0.7828); rgb(12pt)=(0.2645,0.1972,0.7937); rgb(13pt)=(0.2661,0.2011,0.8043); rgb(14pt)=(0.2676,0.2052,0.8148); rgb(15pt)=(0.2691,0.2094,0.8249); rgb(16pt)=(0.2704,0.2138,0.8346); rgb(17pt)=(0.2717,0.2184,0.8439); rgb(18pt)=(0.2729,0.2231,0.8528); rgb(19pt)=(0.274,0.228,0.8612); rgb(20pt)=(0.2749,0.233,0.8692); rgb(21pt)=(0.2758,0.2382,0.8767); rgb(22pt)=(0.2766,0.2435,0.884); rgb(23pt)=(0.2774,0.2489,0.8908); rgb(24pt)=(0.2781,0.2543,0.8973); rgb(25pt)=(0.2788,0.2598,0.9035); rgb(26pt)=(0.2794,0.2653,0.9094); rgb(27pt)=(0.2798,0.2708,0.915); rgb(28pt)=(0.2802,0.2764,0.9204); rgb(29pt)=(0.2806,0.2819,0.9255); rgb(30pt)=(0.2809,0.2875,0.9305); rgb(31pt)=(0.2811,0.293,0.9352); rgb(32pt)=(0.2813,0.2985,0.9397); rgb(33pt)=(0.2814,0.304,0.9441); rgb(34pt)=(0.2814,0.3095,0.9483); rgb(35pt)=(0.2813,0.315,0.9524); rgb(36pt)=(0.2811,0.3204,0.9563); rgb(37pt)=(0.2809,0.3259,0.96); rgb(38pt)=(0.2807,0.3313,0.9636); rgb(39pt)=(0.2803,0.3367,0.967); rgb(40pt)=(0.2798,0.3421,0.9702); rgb(41pt)=(0.2791,0.3475,0.9733); rgb(42pt)=(0.2784,0.3529,0.9763); rgb(43pt)=(0.2776,0.3583,0.9791); rgb(44pt)=(0.2766,0.3638,0.9817); rgb(45pt)=(0.2754,0.3693,0.984); rgb(46pt)=(0.2741,0.3748,0.9862); rgb(47pt)=(0.2726,0.3804,0.9881); rgb(48pt)=(0.271,0.386,0.9898); rgb(49pt)=(0.2691,0.3916,0.9912); rgb(50pt)=(0.267,0.3973,0.9924); rgb(51pt)=(0.2647,0.403,0.9935); rgb(52pt)=(0.2621,0.4088,0.9946); rgb(53pt)=(0.2591,0.4145,0.9955); rgb(54pt)=(0.2556,0.4203,0.9965); rgb(55pt)=(0.2517,0.4261,0.9974); rgb(56pt)=(0.2473,0.4319,0.9983); rgb(57pt)=(0.2424,0.4378,0.9991); rgb(58pt)=(0.2369,0.4437,0.9996); rgb(59pt)=(0.2311,0.4497,0.9995); rgb(60pt)=(0.225,0.4559,0.9985); rgb(61pt)=(0.2189,0.462,0.9968); rgb(62pt)=(0.2128,0.4682,0.9948); rgb(63pt)=(0.2066,0.4743,0.9926); rgb(64pt)=(0.2006,0.4803,0.9906); rgb(65pt)=(0.195,0.4861,0.9887); rgb(66pt)=(0.1903,0.4919,0.9867); rgb(67pt)=(0.1869,0.4975,0.9844); rgb(68pt)=(0.1847,0.503,0.9819); rgb(69pt)=(0.1831,0.5084,0.9793); rgb(70pt)=(0.1818,0.5138,0.9766); rgb(71pt)=(0.1806,0.5191,0.9738); rgb(72pt)=(0.1795,0.5244,0.9709); rgb(73pt)=(0.1785,0.5296,0.9677); rgb(74pt)=(0.1778,0.5349,0.9641); rgb(75pt)=(0.1773,0.5401,0.9602); rgb(76pt)=(0.1768,0.5452,0.956); rgb(77pt)=(0.1764,0.5504,0.9516); rgb(78pt)=(0.1755,0.5554,0.9473); rgb(79pt)=(0.174,0.5605,0.9432); rgb(80pt)=(0.1716,0.5655,0.9393); rgb(81pt)=(0.1686,0.5705,0.9357); rgb(82pt)=(0.1649,0.5755,0.9323); rgb(83pt)=(0.161,0.5805,0.9289); rgb(84pt)=(0.1573,0.5854,0.9254); rgb(85pt)=(0.154,0.5902,0.9218); rgb(86pt)=(0.1513,0.595,0.9182); rgb(87pt)=(0.1492,0.5997,0.9147); rgb(88pt)=(0.1475,0.6043,0.9113); rgb(89pt)=(0.1461,0.6089,0.908); rgb(90pt)=(0.1446,0.6135,0.905); rgb(91pt)=(0.1429,0.618,0.9022); rgb(92pt)=(0.1408,0.6226,0.8998); rgb(93pt)=(0.1383,0.6272,0.8975); rgb(94pt)=(0.1354,0.6317,0.8953); rgb(95pt)=(0.1321,0.6363,0.8932); rgb(96pt)=(0.1288,0.6408,0.891); rgb(97pt)=(0.1253,0.6453,0.8887); rgb(98pt)=(0.1219,0.6497,0.8862); rgb(99pt)=(0.1185,0.6541,0.8834); rgb(100pt)=(0.1152,0.6584,0.8804); rgb(101pt)=(0.1119,0.6627,0.877); rgb(102pt)=(0.1085,0.6669,0.8734); rgb(103pt)=(0.1048,0.671,0.8695); rgb(104pt)=(0.1009,0.675,0.8653); rgb(105pt)=(0.0964,0.6789,0.8609); rgb(106pt)=(0.0914,0.6828,0.8562); rgb(107pt)=(0.0855,0.6865,0.8513); rgb(108pt)=(0.0789,0.6902,0.8462); rgb(109pt)=(0.0713,0.6938,0.8409); rgb(110pt)=(0.0628,0.6972,0.8355); rgb(111pt)=(0.0535,0.7006,0.8299); rgb(112pt)=(0.0433,0.7039,0.8242); rgb(113pt)=(0.0328,0.7071,0.8183); rgb(114pt)=(0.0234,0.7103,0.8124); rgb(115pt)=(0.0155,0.7133,0.8064); rgb(116pt)=(0.0091,0.7163,0.8003); rgb(117pt)=(0.0046,0.7192,0.7941); rgb(118pt)=(0.0019,0.722,0.7878); rgb(119pt)=(0.0009,0.7248,0.7815); rgb(120pt)=(0.0018,0.7275,0.7752); rgb(121pt)=(0.0046,0.7301,0.7688); rgb(122pt)=(0.0094,0.7327,0.7623); rgb(123pt)=(0.0162,0.7352,0.7558); rgb(124pt)=(0.0253,0.7376,0.7492); rgb(125pt)=(0.0369,0.74,0.7426); rgb(126pt)=(0.0504,0.7423,0.7359); rgb(127pt)=(0.0638,0.7446,0.7292); rgb(128pt)=(0.077,0.7468,0.7224); rgb(129pt)=(0.0899,0.7489,0.7156); rgb(130pt)=(0.1023,0.751,0.7088); rgb(131pt)=(0.1141,0.7531,0.7019); rgb(132pt)=(0.1252,0.7552,0.695); rgb(133pt)=(0.1354,0.7572,0.6881); rgb(134pt)=(0.1448,0.7593,0.6812); rgb(135pt)=(0.1532,0.7614,0.6741); rgb(136pt)=(0.1609,0.7635,0.6671); rgb(137pt)=(0.1678,0.7656,0.6599); rgb(138pt)=(0.1741,0.7678,0.6527); rgb(139pt)=(0.1799,0.7699,0.6454); rgb(140pt)=(0.1853,0.7721,0.6379); rgb(141pt)=(0.1905,0.7743,0.6303); rgb(142pt)=(0.1954,0.7765,0.6225); rgb(143pt)=(0.2003,0.7787,0.6146); rgb(144pt)=(0.2061,0.7808,0.6065); rgb(145pt)=(0.2118,0.7828,0.5983); rgb(146pt)=(0.2178,0.7849,0.5899); rgb(147pt)=(0.2244,0.7869,0.5813); rgb(148pt)=(0.2318,0.7887,0.5725); rgb(149pt)=(0.2401,0.7905,0.5636); rgb(150pt)=(0.2491,0.7922,0.5546); rgb(151pt)=(0.2589,0.7937,0.5454); rgb(152pt)=(0.2695,0.7951,0.536); rgb(153pt)=(0.2809,0.7964,0.5266); rgb(154pt)=(0.2929,0.7975,0.517); rgb(155pt)=(0.3052,0.7985,0.5074); rgb(156pt)=(0.3176,0.7994,0.4975); rgb(157pt)=(0.3301,0.8002,0.4876); rgb(158pt)=(0.3424,0.8009,0.4774); rgb(159pt)=(0.3548,0.8016,0.4669); rgb(160pt)=(0.3671,0.8021,0.4563); rgb(161pt)=(0.3795,0.8026,0.4454); rgb(162pt)=(0.3921,0.8029,0.4344); rgb(163pt)=(0.405,0.8031,0.4233); rgb(164pt)=(0.4184,0.803,0.4122); rgb(165pt)=(0.4322,0.8028,0.4013); rgb(166pt)=(0.4463,0.8024,0.3904); rgb(167pt)=(0.4608,0.8018,0.3797); rgb(168pt)=(0.4753,0.8011,0.3691); rgb(169pt)=(0.4899,0.8002,0.3586); rgb(170pt)=(0.5044,0.7993,0.348); rgb(171pt)=(0.5187,0.7982,0.3374); rgb(172pt)=(0.5329,0.797,0.3267); rgb(173pt)=(0.547,0.7957,0.3159); rgb(175pt)=(0.5748,0.7929,0.2941); rgb(176pt)=(0.5886,0.7913,0.2833); rgb(177pt)=(0.6024,0.7896,0.2726); rgb(178pt)=(0.6161,0.7878,0.2622); rgb(179pt)=(0.6297,0.7859,0.2521); rgb(180pt)=(0.6433,0.7839,0.2423); rgb(181pt)=(0.6567,0.7818,0.2329); rgb(182pt)=(0.6701,0.7796,0.2239); rgb(183pt)=(0.6833,0.7773,0.2155); rgb(184pt)=(0.6963,0.775,0.2075); rgb(185pt)=(0.7091,0.7727,0.1998); rgb(186pt)=(0.7218,0.7703,0.1924); rgb(187pt)=(0.7344,0.7679,0.1852); rgb(188pt)=(0.7468,0.7654,0.1782); rgb(189pt)=(0.759,0.7629,0.1717); rgb(190pt)=(0.771,0.7604,0.1658); rgb(191pt)=(0.7829,0.7579,0.1608); rgb(192pt)=(0.7945,0.7554,0.157); rgb(193pt)=(0.806,0.7529,0.1546); rgb(194pt)=(0.8172,0.7505,0.1535); rgb(195pt)=(0.8281,0.7481,0.1536); rgb(196pt)=(0.8389,0.7457,0.1546); rgb(197pt)=(0.8495,0.7435,0.1564); rgb(198pt)=(0.86,0.7413,0.1587); rgb(199pt)=(0.8703,0.7392,0.1615); rgb(200pt)=(0.8804,0.7372,0.165); rgb(201pt)=(0.8903,0.7353,0.1695); rgb(202pt)=(0.9,0.7336,0.1749); rgb(203pt)=(0.9093,0.7321,0.1815); rgb(204pt)=(0.9184,0.7308,0.189); rgb(205pt)=(0.9272,0.7298,0.1973); rgb(206pt)=(0.9357,0.729,0.2061); rgb(207pt)=(0.944,0.7285,0.2151); rgb(208pt)=(0.9523,0.7284,0.2237); rgb(209pt)=(0.9606,0.7285,0.2312); rgb(210pt)=(0.9689,0.7292,0.2373); rgb(211pt)=(0.977,0.7304,0.2418); rgb(212pt)=(0.9842,0.733,0.2446); rgb(213pt)=(0.99,0.7365,0.2429); rgb(214pt)=(0.9946,0.7407,0.2394); rgb(215pt)=(0.9966,0.7458,0.2351); rgb(216pt)=(0.9971,0.7513,0.2309); rgb(217pt)=(0.9972,0.7569,0.2267); rgb(218pt)=(0.9971,0.7626,0.2224); rgb(219pt)=(0.9969,0.7683,0.2181); rgb(220pt)=(0.9966,0.774,0.2138); rgb(221pt)=(0.9962,0.7798,0.2095); rgb(222pt)=(0.9957,0.7856,0.2053); rgb(223pt)=(0.9949,0.7915,0.2012); rgb(224pt)=(0.9938,0.7974,0.1974); rgb(225pt)=(0.9923,0.8034,0.1939); rgb(226pt)=(0.9906,0.8095,0.1906); rgb(227pt)=(0.9885,0.8156,0.1875); rgb(228pt)=(0.9861,0.8218,0.1846); rgb(229pt)=(0.9835,0.828,0.1817); rgb(230pt)=(0.9807,0.8342,0.1787); rgb(231pt)=(0.9778,0.8404,0.1757); rgb(232pt)=(0.9748,0.8467,0.1726); rgb(233pt)=(0.972,0.8529,0.1695); rgb(234pt)=(0.9694,0.8591,0.1665); rgb(235pt)=(0.9671,0.8654,0.1636); rgb(236pt)=(0.9651,0.8716,0.1608); rgb(237pt)=(0.9634,0.8778,0.1582); rgb(238pt)=(0.9619,0.884,0.1557); rgb(239pt)=(0.9608,0.8902,0.1532); rgb(240pt)=(0.9601,0.8963,0.1507); rgb(241pt)=(0.9596,0.9023,0.148); rgb(242pt)=(0.9595,0.9084,0.145); rgb(243pt)=(0.9597,0.9143,0.1418); rgb(244pt)=(0.9601,0.9203,0.1382); rgb(245pt)=(0.9608,0.9262,0.1344); rgb(246pt)=(0.9618,0.932,0.1304); rgb(247pt)=(0.9629,0.9379,0.1261); rgb(248pt)=(0.9642,0.9437,0.1216); rgb(249pt)=(0.9657,0.9494,0.1168); rgb(250pt)=(0.9674,0.9552,0.1116); rgb(251pt)=(0.9692,0.9609,0.1061); rgb(252pt)=(0.9711,0.9667,0.1001); rgb(253pt)=(0.973,0.9724,0.0938); rgb(254pt)=(0.9749,0.9782,0.0872); rgb(255pt)=(0.9769,0.9839,0.0805)},
colorbar
]
\addplot [forget plot] graphics [xmin=-501, xmax=499.8, ymin=-14.4807692307692, ymax=14.3653846153846] {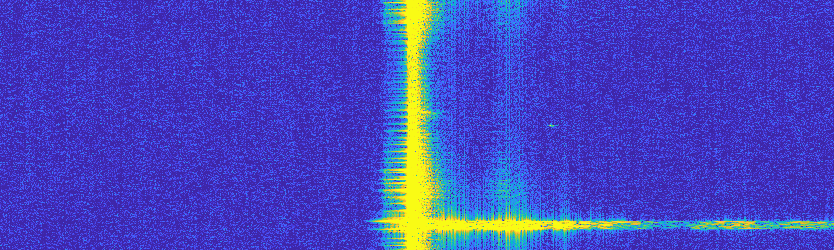};
\end{axis}

\begin{axis}[%
width=1.043\linewidth,
height=0.782\linewidth,
at={(-0.119\linewidth,-0.086\linewidth)},
scale only axis,
point meta min=0,
point meta max=1,
xmin=0,
xmax=1,
ymin=0,
ymax=1,
axis line style={draw=none},
ticks=none,
axis x line*=bottom,
axis y line*=left
]
\end{axis}
\end{tikzpicture}%
         \caption{}
     \end{subfigure}
        \caption{UAV detection on CPI~$3$ with \textbf{(a) No compensation:} Maximum SINR of $16.8$~dB.
                    \textbf{(b) Doppler compensation:} Maximum SINR of $18.7$~dB.
                    \textbf{(c) Stretch compensation:} Maximum SINR of $26.1$~dB.
                    \textbf{(d) Doppler and stretch compensation:} Maximum SINR of $30.4$~dB.
                    }
        \label{fig:SNRComparision}
\end{figure*}

The SINR is investigated with CPIs of $1$~s for five different cases: no compensation, Doppler compensation (Sec.~\ref{sec:DopplerCompensation}), stretch compensation (Sec.~\ref{sec:stretch}), Doppler and stretch compensation, and resampling of the reference waveform (i.e., true stretching). The resampling is performed over the entire CPI, but the CPI is still processed in batches. One could also consider the $2$D correlator, which could potentially further increase the SINR at  significant extra computational cost. The $5$~s data is divided into five CPIs, with the results  presented in Table~\ref{table:SINR}, where the target signal strength is taken to be the strongest resolution cell. The performance of the approximative stretch algorithm is similar to the performance of full resampling. The effect that the different compensations has on the SINR in CPI~$3$ is illustrated in Fig.~\ref{fig:SNRComparision}.

\begin{table}
\centering
\caption{SINR (dB) for different algorithms with $1$~s CPI. The signal strength is taken to be the strongest resolution cell.}
\label{table:SINR}
\begin{tabular}{l|ccccc}
\hline
\hline
Compensation  & CPI 1 & CPI 2 & CPI 3 & CPI 4 & CPI 5 \\
  \hline 
None &   $18.9$  &  $16.2$   &  $16.8$   &   $15.0$  &  $15.5$    \\
Doppler  & $21.0$   & $18.2$  & $18.7$  &  $16.8$  &  $16.6$            \\
Stretch &  $28.2$  & $23.2$ & $26.1$  &  $23.5$ &   $24.7$      \\
Doppler \&  Stretch & $29.7$ & $23.2$  & $30.4$ & $27.1$ & $28.7$                    \\
Doppler \& Resampling & $29.8$ & $23.2$  & $29.5$ & $27.7$  &  $28.1$      \\
\hline
\hline
\end{tabular} 
\end{table}

The SINR processing gain is significant, yielding up to $13.6$~dB improvement for CPI~$3$. Theoretically,  without compensation the signal should spread uniformly in range and Doppler, if the target moves with constant velocity. In our case the spread is not uniform because the signal is relatively weak and effects from interference and noise become noticeable. If instead of the maximum, an average of the smeared target signal is used, the SINR increase is approximately $20$~dB, which is in close agreement to the theoretical loss of $L_D+L_S=2.4~\mathrm{dB} +18.7~\mathrm{dB}= 21.1$~dB, see Fig.~\ref{fig:FBandCorrTimes}. It is not entirely clarified why CPI~$2$ performs  worse than the rest, but a possible explanation is inconsistent flight speed, as the lightweight UAV is  prone to wind disturbances. Indeed, close examination of CPI~$2$ reveals that the UAV accelerates, which is not compensated for by the algorithm.% But it could also be due to other reasons, like scalloping and straddling, etc. 
%Considering that the SINR is not particularly high in the non-compensated case, the exact value is susceptible to noise spikes, meaning the actual SINR increase may be significantly higher. 

\begin{figure}
    \centering
    \begin{subfigure}[b]{0.45\textwidth}
         \centering
        % This file was created by matlab2tikz.
%
%The latest updates can be retrieved from
%  http://www.mathworks.com/matlabcentral/fileexchange/22022-matlab2tikz-matlab2tikz
%where you can also make suggestions and rate matlab2tikz.
%
\definecolor{mycolor1}{rgb}{0.00000,0.44700,0.74100}%
\definecolor{mycolor2}{rgb}{0.85000,0.32500,0.09800}%
\definecolor{mycolor3}{rgb}{0.92900,0.69400,0.12500}%
\definecolor{mycolor4}{rgb}{0.49400,0.18400,0.55600}%
\definecolor{mycolor5}{rgb}{0.46600,0.67400,0.18800}%
\definecolor{mycolor6}{rgb}{0.0,0.0,0.0}%
\begin{tikzpicture}

\begin{axis}[%
width=0.808\linewidth,
height=0.637\linewidth,
at={(0\linewidth,0\linewidth)},
scale only axis,
xmin=-4,
xmax=1.4,
xtick={-3.64385618977472,-2.64385618977472,-1.64385618977472,-0.643856189774725,0.356143810225275,1.35614381022528},
xticklabels={{80},{160},{320},{640},{1280},{2560}},
xlabel style={font=\color{white!15!black}},
xlabel={Integration Time [ms]},
ymin=-2,
ymax=18,
ylabel style={font=\color{white!15!black}},
ylabel={SINR Increase [dB]},
axis background/.style={fill=white},
minor tick num=4,
xminorgrids,
xmajorgrids,
ymajorgrids,
yminorgrids,
legend style={at={(0.02,0.50)}, anchor=south west, legend cell align=left, align=left, draw=white!15!black, font = \footnotesize}
]

\addplot [color=mycolor6, line width=2.0pt, dashed]
  table[row sep=crcr]{%
-3.64385618977473	0\\
-2.64385618977473	3\\
-1.64385618977473	6\\
-0.643856189774725	9\\
0.356143810225275	12\\
1.35614381022527	15\\
};
\addlegendentry{$\text{Theory}$}

\addplot [color=mycolor1, line width=2.0pt]
  table[row sep=crcr]{%
-3.64385618977473	0\\
-2.64385618977473	3.539849822671\\
-1.64385618977473	4.92902138997735\\
-0.643856189774725	9.1241293277787\\
0.356143810225275	8.27507926937708\\
1.35614381022527	6.87373959982545\\
};
\addlegendentry{$\text{T}_{\text{start}}\text{: 0 s}$}

\addplot [color=mycolor2, line width=2.0pt]
  table[row sep=crcr]{%
-3.64385618977473	0\\
-2.64385618977473	-0.564788250950087\\
-1.64385618977473	3.51392010293175\\
-0.643856189774725	4.39011665142363\\
0.356143810225275	3.8717524561747\\
1.35614381022527	10.6807061143975\\
};
\addlegendentry{$\text{T}_{\text{start}}\text{: 0.5 s}$}

\addplot [color=mycolor3, line width=2.0pt]
  table[row sep=crcr]{%
-3.64385618977473	0\\
-2.64385618977473	1.43252721686942\\
-1.64385618977473	1.12534599461399\\
-0.643856189774725	4.04410945314643\\
0.356143810225275	10.6182834641788\\
1.35614381022527	15.3508763968507\\
};
\addlegendentry{$\text{T}_{\text{start}}\text{: 1 s}$}

\addplot [color=mycolor4, line width=2.0pt]
  table[row sep=crcr]{%
-3.64385618977472	0\\
-2.64385618977472	4.42267738166515\\
-1.64385618977472	8.25233279719182\\
-0.643856189774723	12.0006388101602\\
0.356143810225277	15.0674941588863\\
1.35614381022528	16.4304857728834\\
};
\addlegendentry{$\text{T}_{\text{start}}\text{: 1.5 s}$}

\addplot [color=mycolor5, line width=2.0pt]
  table[row sep=crcr]{%
-3.64385618977473	0\\
-2.64385618977473	2.6730757756271\\
-1.64385618977473	5.85306981034016\\
-0.643856189774725	8.44648792523517\\
0.356143810225275	12.2143217585347\\
1.35614381022527	12.0431673362557\\
};
\addlegendentry{$\text{T}_{\text{start}}\text{: 2 s}$}

\end{axis}

\begin{axis}[%
width=1.043\linewidth,
height=0.782\linewidth,
at={(-0.136\linewidth,-0.086\linewidth)},
scale only axis,
xmin=0,
xmax=1,
ymin=0,
ymax=1,
axis line style={draw=none},
ticks=none,
axis x line*=bottom,
axis y line*=left
]
\end{axis}
\end{tikzpicture}%
    \end{subfigure}
    \caption{SINR gain as a function of integration time for different sequences of the $5$~s recorded data. The dashed black line illustrates the expected SINR increase of $3$~dB increase per doubling of the integration time. The improvement is occasionally better than this, which can be explained by changes in straddling, aspect angle, multipath, etc.}
        \label{fig:SNRvsInt}
\end{figure}
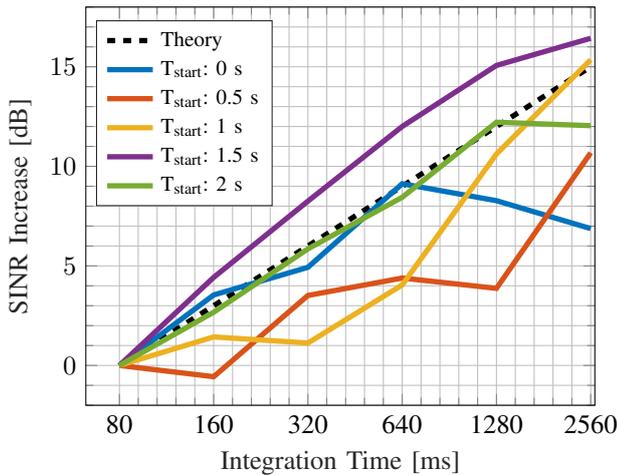

When compensating for Doppler and cell migration, SINR gain as a function of integration time is investigated to estimate the coherence time of the UAV. The results are shown in Fig.~\ref{fig:SNRvsInt}, where a coherent integration time of up to $2.5$~s is observed to be possible. We expect that several factors beside acceleration, such as multipath propagation and changes in aspect angle, impact the coherence time of the target.% \textcolor{red}{It can also explain why the SINR occasionally increases more than the $3$~dB per doubling in integration time.}
%It is not straight forward to classify a coherent integration time, it seems to very much depend on environmental factors (unsurprisingly). Considering acceleration is not accounted for, fluctuations in velocity will lead to a loss. %\textcolor{red}{How much effect this has likely depends on the timescale, for short timescales the radar can not perceive the fluctuations, for long timescales the average will be taken and it probably does not matter, for medium however it can probably have hugh effect.  } 
%Also, the batched implementation suffers from scalloping and straddling losses. %\textcolor{red}{ which likely can be mitigated by oversampling, windowing, zero padding, etc.} 
%Still, the coherent integration time, in certain situations, can be upwards to $2.5$~s.

%xtcolor{blue}{An alternative to coherent integration is incoherent integration. For example, coherent integration can be performed for CPIs of $XXX$~ms, whereas several CPIs are then incoherently integrated, i.e., simply summation of the different CPIS. .... om plats}

\section{Conclusion}
\label{sec:Conclusions}
In summary, we have experimentally investigated moving target compensation for high time-bandwidth noise radars and shown that significant increase in SINR can be achieved by compensating for the Doppler shift and the target's cell migration. The coherence time of the target UAV fluctuates considerably, but  times of at least $2.5$~s are observed for a subset of the data.   The compensation is performed using an approximative stretch algorithm suitable for real-time implementation. In future work, this algorithm will be implemented onto a real-time high time-bandwidth noise radar processor.

\section*{Acknowledgement}
The authors acknowledge support from the Knut and Alice Wallenberg (KAW) Foundation through the Wallenberg Centre for Quantum Technology
(WACQT). %\textcolor{blue}{The authors would like to thank our fellow colleagues for support and rewarding  discussions. } 
%Kent Falk -rent av medförfattare kanske
\bibliography{bib.bib}

% Generated by IEEEtran.bst, version: 1.14 (2015/08/26)
\begin{thebibliography}{10}
\providecommand{\url}[1]{#1}
\csname url@samestyle\endcsname
\providecommand{\newblock}{\relax}
\providecommand{\bibinfo}[2]{#2}
\providecommand{\BIBentrySTDinterwordspacing}{\spaceskip=0pt\relax}
\providecommand{\BIBentryALTinterwordstretchfactor}{4}
\providecommand{\BIBentryALTinterwordspacing}{\spaceskip=\fontdimen2\font plus
\BIBentryALTinterwordstretchfactor\fontdimen3\font minus \fontdimen4\font\relax}
\providecommand{\BIBforeignlanguage}[2]{{%
\expandafter\ifx\csname l@#1\endcsname\relax
\typeout{** WARNING: IEEEtran.bst: No hyphenation pattern has been}%
\typeout{** loaded for the language `#1'. Using the pattern for}%
\typeout{** the default language instead.}%
\else
\language=\csname l@#1\endcsname
\fi
#2}}
\providecommand{\BIBdecl}{\relax}
\BIBdecl

\bibitem{LPIRadarStrategies}
A.~G. Stove, A.~L. Hume, and C.~J. Baker, ``Low probability of intercept radar strategies,'' \emph{IEE Proceedings - Radar, Sonar and Navigation}, vol. 151, pp. 249--260, 2004.

\bibitem{pace2009detecting}
P.~E. Pace, \emph{Detecting and classifying low probability of intercept radar}, 2nd~ed.\hskip 1em plus 0.5em minus 0.4em\relax Norwood, MA, USA: Artech house, 2009.

\bibitem{Horton}
B.~M. Horton, ``Noise-modulated distance measuring systems,'' \emph{Proceedings of the IRE}, vol.~47, pp. 821--828, 1959.

\bibitem{GeneralizedAmbiguityfunction}
M.~Dawood and R.~M. Narayanan, ``Generalised wideband ambiguity function of a coherent ultrawideband random noise radar,'' \emph{IEE Proceedings - Radar, Sonar, and Navigation}, vol. 150, pp. 379--386, 2003.

\bibitem{SuneAmbiguity}
S.~R.~J. Axelsson, ``Noise radar using random phase and frequency modulation,'' \emph{IEEE Transactions on Geoscience and Remote Sensing}, vol.~42, pp. 2370 -- 2384, 2004.

\bibitem{PulseShaping}
F.~De~Palo \emph{et~al.}, ``Introduction to noise radar and its waveforms,'' \emph{Sensors}, vol.~20, 2020.

\bibitem{Stretch:TimeTranformation}
W.~J. Caputi, ``Stretch: A time-transformation technique,'' \emph{IEEE Transactions on Aerospace and Electronic Systems}, vol. AES-7, pp. 269--278, 1971.

\bibitem{FastPulseDopplerRadarProcessing}
T.~L. Marzetta, E.~A. Martinsen, and C.~P. Plum, ``Fast pulse {D}oppler radar processing accounting for range bin migration,'' in \emph{The Record of the 1993 IEEE National Radar Conference}.\hskip 1em plus 0.5em minus 0.4em\relax Lynnfield, USA: IEEE, 1993, pp. 264--268.

\bibitem{StretchProcessingLongIntTime}
K.~S. Kulpa and J.~Misiurewicz, ``Stretch processing for long integration time passive covert radar,'' in \emph{2006 CIE International Conference on Radar}.\hskip 1em plus 0.5em minus 0.4em\relax Shanghai, China: IEEE, 2006, pp. 1--4.

\bibitem{xu2011radon}
J.~Xu \emph{et~al.}, ``Radon-{F}ourier transform for radar target detection, {I}: {G}eneralized {D}oppler filter bank,'' \emph{IEEE transactions on aerospace and electronic systems}, vol.~47, pp. 1186--1202, 2011.

\bibitem{Shan2016}
T.~Shan \emph{et~al.}, ``Efficient architecture and hardware implementation of coherent integration processor for digital video broadcast-based passive bistatic radar,'' \emph{IET Radar, Sonar \& Navigation}, vol.~10, pp. 97--106, 2016.

\bibitem{EffectsofMovementforHighTimeBandwidths}
D.~Bok, D.~O’Hagan, and P.~Knott, ``Effects of movement for high time-bandwidths in batched pulse compression range-{D}oppler radar,'' \emph{Sensors}, vol.~21, 2021.

\bibitem{DopplerTolerance}
C.~Wasserzier, ``Exploiting the low {D}oppler tolerance of noise radar to perform precise velocity measurements on a short set of data,'' \emph{Signals}, vol.~2, pp. 25--40, 2021.

\bibitem{BatchProcessing2}
S.~R.~J. Axelsson, ``Noise radar for range/{D}oppler processing and digital beamforming using low-bit {ADC},'' \emph{IEEE Transactions on Geoscience and Remote Sensing}, vol.~41, pp. 2703--2720, 2003.

\bibitem{UlanderFactorizedBackProjection}
L.~M.~H. Ulander, H.~Hellsten, and G.~Stenstr\"om, ``Synthetic-aperture radar processing using fast factorized back-projection,'' \emph{IEEE Transactions on Aerospace and Electronic Systems}, vol.~39, pp. 760--776, 2003.

\bibitem{Vivace}
\BIBentryALTinterwordspacing
{Intermodulation Products AB}, \emph{{Intermodulation Products – Vivace (2021)}}, Segersta, Sweden, Accessed: May 17, 2023. [Online]. Available: \url{https://intermod.pro/products/microwave-platforms}
\BIBentrySTDinterwordspacing

\bibitem{DemonstrationofCorrelationNoiseSuppression}
M.~Ankel \emph{et~al.}, ``Bistatic noise radar: {D}emonstration of correlation noise suppression,'' \emph{IET Radar, Sonar \& Navigation}, vol.~17, pp. 351--361, 2023.

\bibitem{MaskingeffectSune}
S.~R.~J. Axelsson, ``Analysis of random step frequency radar and comparison with experiments,'' \emph{IEEE Transactions on Geoscience and Remote Sensing}, vol.~45, pp. 890--904, 2007.

\end{thebibliography}

%\newpage 
\end{document}